\documentstyle[12pt]{article}
\begin{document}
\begin{flushright}
hep-th/0507127\\ SNB-BHU/Preprint
\end{flushright}
\vskip 1.2cm
\begin{center}
{\bf \Large { Unique nilpotent symmetry transformations for \\
matter fields in QED: augmented superfield formalism }}

\vskip 1.5cm

{\bf R.P.Malik}\\
{\it S. N. Bose National Centre for Basic Sciences,} \\
{\it Block-JD, Sector-III, Salt Lake, Calcutta-700 098, India} \\

\vskip .2cm

and\\

\vskip .2cm

{\it Centre of Advanced Studies, Physics Department,}\\
{\it Banaras Hindu University, Varanasi-221 005, India}\\
{\bf E-mail address: malik@bhu.ac.in}

\vskip 1.5cm

\end{center}

\noindent {\bf Abstract}: We derive the off-shell nilpotent
(anti-)BRST symmetry transformations for the interacting $U(1)$
gauge theory of quantum electrodynamics (QED) in the framework of
augmented superfield approach to the BRST formalism. In addition
to the horizontality condition, we invoke another gauge invariant
condition on the six $(4, 2)$-dimensional supermanifold to obtain
the exact and unique nilpotent symmetry transformations for {\it
all} the basic fields, present in the (anti-)BRST invariant
Lagrangian density of the physical four $(3 + 1)$-dimensional QED.
The above supermanifold is parametrized by four even spacetime
variables $x^\mu$ (with $\mu = 0, 1, 2, 3$) and a couple of odd
variables ($\theta$ and $\bar\theta$) of the Grassmann algebra.
The new gauge invariant condition on the supermanifold owes its
origin to the (super) covariant derivatives and leads to the
derivation of unique nilpotent symmetry transformations for the
matter fields. The geometrical interpretations for all the above
off-shell nilpotent (anti-)BRST transformations are discussed,
too. \baselineskip=16pt

\vskip 0.7cm



\newpage

\noindent
{\bf 1 Introduction}\\

\noindent The usual superfield approach [1-5] to
Becchi-Rouet-Stora-Tyutin (BRST) formalism (see, e.g., [6-9] for
details) for a $p$-form (with $p = 1, 2, 3....)$ Abelian gauge
theory delves deep into the geometrical aspects of the nilpotent
(anti-)BRST symmetries (and corresponding nilpotent generators)
for the $p$-form gauge fields and the underlying (anti-)ghost
fields of the theory. To be precise, under the above approach, the
$D$-dimensional gauge theory is first considered on the $(D,
2)$-dimensional supermanifold which is parametrized by the
$D$-number of even spacetime commuting coordinates $x^\mu$ (with
$\mu = 0, 1, 2.....D-1$) and two anticommuting (i.e. $\theta^2 =
\bar \theta^2 = 0, \theta \bar\theta + \bar \theta \theta = 0$)
odd variables ($\theta$ and $\bar\theta$) of the Grassmann
algebra. After this, a $(p + 1)$-form super curvature $\tilde
F^{(p + 1)} = \tilde d \tilde A^{(p)}$ is constructed from (i) the
super exterior derivative $\tilde d = dx^\mu \partial_\mu +
d\theta \partial_\theta + d \bar\theta \partial_{\bar\theta} $
(with $\tilde d^2 = 0$), and (ii) the super $p$-form connection
$\tilde A^{(p)}$ on the $(D, 2)$-dimensional supermanifold.
Subsequently, this super curvature $\tilde F^{(p + 1)}$ is
equated, due to the so-called horizontality condition [1-5], with
the ordinary $(p + 1)$-form curvature $ F^{(p + 1)} =  d  A^{(p)}
$ constructed by the ordinary $D$-dimensional exterior derivative
$d = dx^\mu \partial_\mu$ (with $d^2 = 0$) and the ordinary
$p$-form connection $A^{(p)}$ defined on the ordinary
$D$-dimensional Minkowskian flat spacetime manifold on which the
starting $p$-form gauge theory (endowed with the first-class
constraints) exists.

The above horizontality condition is christened as the
soul-flatness condition in [6] which mathematically amounts to
setting equal to zero all the Grassmannian components of the
(anti-) symmetric tensor that defines the $(p + 1)$-form super
curvature $\tilde F^{(p + 1)}$ on the $(D, 2)$-dimensional
supermanifold. The process of reduction of the $(D,
2)$-dimensional super curvature to the $D$-dimensional ordinary
curvature (i.e. the equality $\tilde F^{(p + 1)} = F^{(p + 1)}$)
leads to the derivation of the nilpotent (anti-)BRST symmetry
transformation for the $p$-form gauge fields and the (anti-)
commuting (anti-) ghost fields of the theory \footnote{It can be
seen that, for the 2-form Abelian gauge theory, the bosonic and
fermionic ghosts do exist in the BRST formalism and their
transformations can be derived using the usual superfield
formalism [10].}. As a bonus and by-product, the geometrical
interpretations for the nilpotency and anticommutativity
properties of the conserved and nilpotent (anti-)BRST charges
\footnote{These charges turn out to be the translational
generators along the Grassmannian directions of the supermanifold.
Their nilpotency and anticommutativity properties are also found
to be encoded in the specific properties associated with the above
translational generators (see, e.g., [11-16] for details).} emerge
very naturally on the supermanifold. However, these beautiful
connections between the geometrical aspects of the supermanifold
and the (anti-)BRST transformations (and their corresponding
generators) remain absolutely confined to the gauge fields and the
(anti-)ghost fields. The above usual superfield formalism [1-5,
10] does not shed any light on the (anti-)BRST symmetry
transformations associated with the matter fields of an
interacting Abelian gauge theory.

It is worthwhile to mention, at this juncture, that the usual
superfield approach has also been applied to the case of four $(3
+ 1)$-dimensional (4D) 1-form ($A^{(1)} = dx^\mu A_\mu$)
non-Abelian gauge theory where the 2-form super curvature $\tilde
F^{(2)} = \tilde d \tilde A^{(1)} + \tilde A^{(1)} \wedge \tilde
A^{(1)}$, defined on the six (4, 2)-dimensional supermnifold,  is
equated to the 4D ordinary 2-form curvature $F^{(2)} = d A^{(1)} +
A^{(1)} \wedge A^{(1)}$ (constructed from $d = dx^\mu
\partial_\mu$ and $A^{(1)}$) due to the horizontality condition.
As expected, this procedure of covariant reduction of the 2-form
super curvature to ordinary curvature leads to the derivation of
nilpotent (anti-)BRST symmetry transformations, associated with
the non-Abelian gauge field and the anticommuting (anti-)ghost
fields of the theory (see, e.g., [4]). The matter (Dirac) fields
of the interacting non-Abelian gauge theory are found to play no
role in the above covariant reduction associated with the
horizontality condition. As a consequence, one does not obtain the
nilpotent (anti-)BRST symmetry transformations for the matter
(Dirac) fields by the usual superfield formalism.

The purpose of our present paper is to derive uniquely and exactly
the off-shell nilpotent (anti-)BRST symmetry transformations for
the matter (Dirac) fields of QED in 4D by invoking a new gauge
invariant restriction, besides the usual horizontality condition,
on the six (4, 2)-dimensional supermanifold. In this context, it
is worthwhile to point out that, in a recent set of papers
[11-16], the usual superfield approach (with horizontality
condition {\it alone}) has been extended to include the invariance
of the conserved (matter) currents/charges to obtain all the
nilpotent  and anticommuting (anti-)BRST symmetry transformations
for {\it all} the basic fields of the interacting (non-)Abelian
gauge theories as well as the reparametrization invariant free
scalar and spinning relativistic particle(s). However, the
nilpotent (anti-)BRST symmetry transformations, that emerge due to
the latter restrictions on the supermanifold, are {\it not} proved
to be mathematically {\it unique}. One of the central themes of
our present paper is to demonstrate that the new gauge invariant
restriction on the supermanifold (cf. (4.1)), that owes its origin
to the (super) covariant derivatives, leads to the derivation of
the off-shell nilpotent (anti-)BRST symmetry transformations for
the matter fields uniquely. It is very gratifying that the
geometrical interpretations for the (anti-)BRST symmetry
transformations and their corresponding nilpotent generators (that
emerge especially after the application of the horizontality
condition) remain intact under the above extended version of the
usual superfield approach to BRST formalism. Thus, there is very
neat mutual consistency, conformity and complementarity  between
the above two restrictions on the supermanifold. We christen our
present approach as well as that of [11-16] as the augmented
superfield formalism because (i) all these attempts are a set of
consistent extensions (and, in some sense, generalizations) of the
usual superfield approach to BRST formalism, and (ii) the
nilpotent and anticommuting (anti-)BRST transformations for {\it
all} the fields of the 4D interacting 1-form Abelian $U(1)$ gauge
theory are derived in this superfield approach to BRST formalism.

The contents of our present paper are organized as follows. In
Sec. 2, we recapitulate the bare essentials of the off-shell
nilpotent (anti-)BRST symmetry transformations in the Lagrangian
formulation for the 4D {\it interacting} $U(1)$ gauge theory
(QED). For the paper to be self-contained, Sec. 3 is devoted to a
brief description of the derivation of the above symmetry
transformations for the gauge field $A_\mu$ and the (anti-) ghost
fields $(\bar C)C$ by exploiting the usual horizontality condition
on the six $(4, 2)$-dimensional
 supermanifold [4,11,12]. The central result of our paper is contained
in Sec. 4  where we derive uniquely the off-shell nilpotent
symmetry transformations for the matter (Dirac) fields in the framework
of the augmented superfield formalism by exploiting a gauge invariant
restriction on the six (4, 2)-dimensional supermanifold. Its gauge covariant
version does not lead to the derivation of correct nilpotent symmetries
(see, e.g., Appendix). Finally, we make some concluding remarks
and pinpoint a few future directions for further investigations in Sec. 5.\\

\noindent {\bf 2 Preliminary: nilpotent (anti-)BRST symmetries}\\

\noindent To  set the notations and conventions that will be
useful for our later discussions, we begin with the (anti-)BRST
invariant Lagrangian density ${\cal L}_B$ for the interacting four
$(3 + 1)$-dimensional (4D) $U(1)$ gauge theory (i.e. QED) in the
Feynman gauge \footnote{We adopt here the conventions and
notations such that the 4D flat Minkowski metric: $\eta_{\mu\nu}
=$ diag $(+1, -1, - 1, -1)$ and $\Box = \eta^{\mu\nu}
\partial_{\mu} \partial_{\nu} = (\partial_{0})^2 -
(\partial_{1})^2 - (\partial_2)^2 - (\partial_3)^2, F_{0i} = E_i =
\partial_{0} A_{i} - \partial_{i} A_{0} =  F^{i0}, F_{ij} = -
\varepsilon_{ijk}\; B_k, B_i = - \frac{1}{2} \varepsilon_{ijk}
F_{jk},
\partial^\mu A_{\mu} = \partial_\mu A^\mu = (\partial \cdot A)
\equiv \partial_0 A_0 - \partial_i A_i$. Here $E_i$ and $B_i$ are the
electric and magnetic fields and $\varepsilon_{ijk}$ is the
3D totally antisymmetric
Levi-Civita tensor in the space indices.
Furthermore, the Greek indices $\mu, \nu, \rho...
= 0, 1, 2, 3$ correspond to the spacetime directions on the 4D
Minkowskian spacetime manifold and the Latin indices $i, j, k...= 1 , 2, 3$
stand for the space directions only.} [6-9]:
$$
\begin{array}{lcl}
{\cal L}_{B} &=& - \frac{1}{4}\; F^{\mu\nu} F_{\mu\nu}
+ \bar \psi \;(i \gamma^\mu D_\mu - m)\; \psi + B \;(\partial \cdot A)
+ \frac{1}{2}\; B^2
- i \;\partial_{\mu} \bar C \partial^\mu C,
\end{array} \eqno(2.1)
$$
where $F_{\mu\nu} = \partial_\mu A_\nu - \partial_\nu A_\mu$ is the field
strength tensor for the $U(1)$ gauge theory and the covariant derivative on
the matter (Dirac) field
$D_\mu \psi = \partial_\mu \psi + i e A_\mu \psi$ leads to the
interaction term between the $U(1)$ gauge field $A_\mu$ and
Dirac fields $\psi$ and $\bar\psi$ of mass $m$ and electric charge $e$
(i.e. $- e \bar \psi \gamma^\mu A_\mu \psi$). In fact, this
term arises through the term
$i \bar \psi \gamma^\mu D_\mu \psi$ that is
 present in the Lagrangian density (2.1)
 where $\gamma^\mu$'s are the $4 \times 4$ Dirac matrices.
The anticommuting ($ C \bar C + \bar C C = 0, C^2 = \bar C^2 = 0,
C \psi + \psi C = 0$, etc.) (anti-) ghost fields $(\bar C)C$ are
required to maintain the unitarity and ``quantum'' gauge (i.e.
BRST) invariance together at any arbitrary order of perturbation
theory \footnote{ The full strength of the (anti-) ghost fields
turns up in the discussion of the unitarity and gauge invariance
for the perturbative computations in the realm of non-Abelian
gauge theory where the loop diagrams of the gauge (gluon) fields
play a very important role. In fact, for each such a gluon loop
diagram, a ghost loop diagram is required for the precise proof of
unitarity in the theory (see, e.g., [17,7]).}. The
Nakanishi-Lautrup auxiliary  field $B$ is required to linearize
the gauge-fixing term $-\frac{1}{2} (\partial\cdot A)^2$ in (2.1).
The above Lagrangian density (2.2) respects the following local,
covariant, continuous, off-shell nilpotent $(s_{(a)b}^2 = 0)$ and
anticommuting ($s_b s_{ab} + s_{ab} s_b = 0$) (anti-)BRST
($s_{(a)b}$) \footnote{We adopt here the notations and conventions
followed in [8,9]. In fact, in its full blaze of glory, a
nilpotent ($\delta_{B}^2 = 0$) BRST transformation $\delta_{B}$ is
equal to the product of an anticommuting ($\eta C = - C \eta, \eta
\bar C = - \bar C\eta, \eta \psi = - \psi \eta, \eta \bar \psi = -
\bar \psi \eta$, etc.) spacetime independent parameter $\eta$ and
$s_{b}$ (i.e. $\delta_{B} = \eta \; s_{b}$) with $s_{b}^2 = 0$.}
symmetry transformations (see, e.g., [6-9] for all the details) $$
\begin{array}{lcl}
s_{b} A_{\mu} &=& \partial_{\mu} C, \qquad
s_{b} C = 0, \qquad
s_{b} \bar C = i B,  \qquad s_b \psi = - i e C \psi, \nonumber\\
s_b \bar \psi &=& - i e \bar \psi C,
\qquad  s_{b} B = 0, \quad
\;s_{b} F_{\mu\nu} = 0, \quad s_b (\partial \cdot A) = \Box C, \nonumber\\
s_{ab} A_{\mu} &=& \partial_{\mu} \bar C, \qquad
s_{ab} \bar C = 0, \qquad
s_{ab} C = - i B,  \qquad s_{ab} \psi = - i e \bar C \psi, \nonumber\\
s_{ab} \bar \psi &=& - i e \bar \psi \bar C,
\qquad  s_{ab} B = 0, \quad
\;s_{ab} F_{\mu\nu} = 0, \quad s_{ab} (\partial \cdot A) = \Box \bar C.
\end{array}\eqno(2.2)
$$ The noteworthy points, at this stage, are (i) under the
(anti-)BRST transformations, it is the kinetic energy term (more
precisely $F_{\mu\nu}$ itself) that remains invariant and
gauge-fixing term $(\partial \cdot A)$ transforms. It should be
emphasized that the antisymmetric field strength tensor
$F_{\mu\nu}$ remains invariant under the original local gauge
transformation (i.e. $\delta_g A_\mu = \partial_\mu \alpha (x)$
with $\alpha (x)$ as an infinitesimal gauge parameter), too. In
fact, all the gauge invariant quantities remain invariant
quantities under the (anti-)BRST transformations as well. (ii) The
starting local $U(1)$ gauge invariant theory is endowed with the
first-class constraints in the language of Dirac's classification
scheme for constraints. These constraints are found to be encoded
in the physicality criteria where physical states ($|phys>$) (of
the total Hilbert space of quantum states) are annihilated (i.e.
$Q_{(a)b} |phys> = 0$) by the conserved and nilpotent (anti-)BRST
charges $Q_{(a)b}$ (see, e.g., [6-9] for details). (iii) The
local, conserved and nilpotent charges $Q_{(a)b}$ can be computed
by exploiting the Noether theorem. These charges do generate the
nilpotent and anticommuting (anti-)BRST transformations. This
statement can be succinctly expressed in the mathematical form as
given below $$
\begin{array}{lcl}
s_{r}\; \Omega (x) = - i\;
\bigl [\; \Omega (x),  Q_r\; \bigr ]_{\pm}, \qquad\;\;\;
r = b, ab,
\end{array} \eqno(2.3)
$$ where the local generic field $\Omega = A_\mu, C, \bar C, \psi,
\bar \psi, B$ and the $(+)-$ signs, as the subscripts on the
square bracket, stand for the bracket to be an (anti-)commutator
for the local generic field $\Omega$ being (fermionic) bosonic in
nature.\\

\noindent {\bf 3 Symmetries for gauge- and (anti-)ghost fields:
usual superfield formalism}\\

\noindent To obtain the off-shell nilpotent symmetry
transformations for the gauge field $A_\mu$ and the (anti-)ghost
fields $(\bar C)C$, we begin with a six ($4, 2$)-dimensional
supermanifold parametrized by the superspace coordinates $Z^M =
(x^\mu, \theta, \bar \theta)$ where $x^\mu\; (\mu = 0, 1, 2, 3)$
are the even (bosonic) spacetime coordinates and $\theta$ and
$\bar \theta$ are the two odd (Grassmannian) coordinates. On this
supermanifold, one can define a 1-form super connection $\tilde
A^{(1)} = dZ^M \tilde A_M$ where $\tilde A_M = (B_{\mu} (x,
\theta, \bar \theta), \;{\cal F} (x, \theta, \bar \theta), \;{\bar
{\cal F}} (x, \theta, \bar \theta))$ are the component multiplet
superfields [4, 3] with $B_\mu$ being bosonic in nature and ${\cal
F}, \bar {\cal F}$ being fermionic (i.e. ${\cal F}^2 = \bar {\cal
F}^2 = 0$). The superfields $B_{\mu} (x,\theta,\bar\theta), {\cal
F} (x,\theta,\bar\theta), \bar {\cal F} (x,\theta,\bar\theta)$ can
be expanded in terms of the basic fields $A_\mu (x), C (x), \bar C
(x)$, auxiliary field $B (x)$ of  (2.1) and some extra fields as
[4,3,11] $$
\begin{array}{lcl}
B_{\mu} (x, \theta, \bar \theta) &=& A_{\mu} (x) + \theta\; \bar
R_{\mu} (x) + \bar \theta\; R_{\mu} (x) + i \;\theta \;\bar \theta
S_{\mu} (x), \nonumber\\ {\cal F} (x, \theta, \bar \theta) &=& C
(x) + i\; \theta \bar B (x) + i \;\bar \theta\; {\cal B} (x) + i\;
\theta\; \bar \theta \;s (x), \nonumber\\ \bar {\cal F} (x,
\theta, \bar \theta) &=& \bar C (x) + i \;\theta\;\bar {\cal B}
(x) + i\; \bar \theta \;B (x) + i \;\theta \;\bar \theta \;\bar s
(x).
\end{array} \eqno(3.1)
$$
It is straightforward to note that the local
fields $ R_{\mu} (x), \bar R_{\mu} (x),
C (x), \bar C (x), s (x), \bar s (x)$ are fermionic (anticommuting)
in nature and $A_{\mu} (x), S_{\mu} (x), {\cal B} (x), \bar {\cal B} (x),
B (x), \bar B (x)$ are bosonic (i.e. commuting) so that, in the
above expansion, the bosonic-
 and fermionic degrees of freedom match. This requirement is essential
for the sanctity of any arbitrary supersymmetric theory described in the
framework of
superfield formulation. In fact, all the secondary fields will be expressed
in terms of basic fields (and their derivatives)
due to the restrictions emerging from the application
of horizontality condition, namely;
$$
\begin{array}{lcl}
\frac{1}{2}\; (d Z^M \wedge d Z^N)\;
\tilde F_{MN} = \tilde d \tilde A^{(1)}  \equiv
d A^{(1)} = \frac{1}{2} (dx^\mu \wedge dx^\nu)\; F_{\mu\nu},
\end{array} \eqno(3.2)
$$
where the super exterior derivative $\tilde d$ and
the connection super one-form $\tilde A^{(1)}$ are defined as
$$
\begin{array}{lcl}
\tilde d &=& \;d Z^M \;\partial_{M} = d x^\mu\; \partial_\mu\;
+ \;d \theta \;\partial_{\theta}\; + \;d \bar \theta \;\partial_{\bar \theta},
\nonumber\\
\tilde A^{(1)} &=& d Z^M\; \tilde A_{M} = d x^\mu \;B_{\mu} (x , \theta, \bar \theta)
+ d \theta\; \bar {\cal F} (x, \theta, \bar \theta) + d \bar \theta\;
{\cal F}( x, \theta, \bar \theta).
\end{array}\eqno(3.3)
$$
In physical language, this requirement
(i.e. equation (3.2)) implies that the physical fields
$E_i$ and $B_i$, derived from the curvature term $F_{\mu\nu}$, do not get any
contribution from the Grassmannian variables. In other words, the
physical electric and magnetic fields ($E_i$
and $B_i$ for the 4D QED) remain unchanged in the
superfield formulation. Mathematically, the condition (3.2) implies
the ``flatness'' of all the components of the
super curvature tensor $\tilde F_{MN}$
(derived from the super 2-form) that are directed along the
 $\theta$ and/or $\bar \theta$ directions of the supermanifold. To this
end in mind, let us first expand $\tilde d \tilde A^{(1)}$ explicitly as
$$
\begin{array}{lcl}
\tilde d \tilde A^{(1)} &=& (d x^\mu \wedge d x^\nu)\;
(\partial_{\mu} B_\nu) - (d \theta \wedge d \theta)\; (\partial_{\theta}
\bar {\cal F}) + (d x^\mu \wedge d \bar \theta)
(\partial_{\mu} {\cal F} - \partial_{\bar \theta} B_{\mu}) \nonumber\\
&-& (d \theta \wedge d \bar \theta) (\partial_{\theta} {\cal F}
+ \partial_{\bar \theta} \bar {\cal F})
+ (d x^\mu \wedge d \theta) (\partial_{\mu} \bar {\cal F} - \partial_{\theta}
B_{\mu}) - (d \bar \theta \wedge d \bar \theta)
(\partial_{\bar \theta} {\cal F}).
\end{array}\eqno(3.4)
$$
Ultimately, the application of soul-flatness (horizontality) condition
($\tilde d \tilde A^{(1)} = d A^{(1)}$) leads to the
following relationships between extra secondary fields and basic fields
(and their derivatives) (see, e.g., [11, 12] for all the details)
$$
\begin{array}{lcl}
R_{\mu} \;(x) &=& \partial_{\mu}\; C(x), \qquad
\bar R_{\mu}\; (x) = \partial_{\mu}\;
\bar C (x), \qquad \;s\; (x) = \bar s\; (x) = 0,
\nonumber\\
S_{\mu}\; (x) &=& \partial_{\mu} B\; (x)
\qquad
B\; (x) + \bar B \;(x) = 0, \qquad
{\cal B}\; (x)  = \bar {\cal B} (x) = 0.
\end{array} \eqno(3.5)
$$
The insertion of all the above values in the expansion (3.1) leads to
$$
\begin{array}{lcl}
B^{(h)}_{\mu}\; (x, \theta, \bar \theta) &=& A_{\mu} (x)
+ \;\theta\; \partial_\mu \bar C (x)
+ \;\bar \theta\; \partial_\mu C (x)
+ i \;\theta \;\bar \theta \;\partial_\mu B (x), \nonumber\\
{\cal F}^{(h)}\; (x, \theta, \bar \theta) &=& C (x) \;- i\; \theta\; B (x),
\qquad
\bar {\cal F}^{(h)}\; (x, \theta, \bar \theta) = \bar C (x)
+ i \;\bar \theta\; B (x),
\end{array} \eqno(3.6)
$$
where our starting super expansion for the
multiplet super fields of (3.1) have changed to:
$B_\mu \to B_\mu^{(h)}, {\cal F} \to {\cal F}^{(h)}, \bar {\cal F} \to \bar {\cal F}^{(h)}$
after the application of the horizontality condition. As a result
$\tilde A^{(1)}  \to \tilde A^{(1)}_{(h)} =
 d x^\mu B^{(h)}_{\mu}
+ d \theta \bar {\cal F}^{(h)} + d \bar \theta {\cal F}^{(h)}$ in
(3.3). In fact, the above reduction leads to the derivation of the
(anti-)BRST symmetries for the gauge- and (anti-) ghost fields of
the Abelian gauge theory. In addition, this exercise provides  the
physical interpretation for the (anti-)BRST charges $Q_{(a)b}$ as
the generators (cf. (2.3)) of translations (i.e. $
(\mbox{Lim}_{\bar\theta \rightarrow 0} (\partial/\partial \theta),
 \mbox{Lim}_{\theta \rightarrow 0} (\partial/\partial \bar\theta)$)
along the Grassmannian
directions of the supermanifold. Both these observations can be succinctly
expressed, in a combined way, by re-writing the super expansion (3.6) as
$$
\begin{array}{lcl}
B^{(h)}_{\mu}\; (x, \theta, \bar \theta) &=& A_{\mu} (x)
+ \;\theta\; (s_{ab} A_{\mu} (x))
+ \;\bar \theta\; (s_{b} A_{\mu} (x))
+ \;\theta \;\bar \theta \;(s_{b} s_{ab} A_{\mu} (x)), \nonumber\\
{\cal F}^{(h)}\; (x, \theta, \bar \theta) &=& C (x) \;+ \; \theta\; (s_{ab} C (x))
\;+ \;\bar \theta\; (s_{b} C (x))
\;+ \;\theta \;\bar \theta \;(s_{b}\; s_{ab} C (x)),
 \nonumber\\
\bar {\cal F}^{(h)}\; (x, \theta, \bar \theta) &=& \bar C (x)
\;+ \;\theta\;(s_{ab} \bar C (x)) \;+\bar \theta\; (s_{b} \bar C (x))
\;+\;\theta\;\bar \theta \;(s_{b} \;s_{ab} \bar C (x)),
\end{array} \eqno(3.7)
$$ where, it is evident that, $s_b C = 0$ and $s_{ab} \bar C = 0$
have been taken into account. In fact, it is because of these
inputs that the above expansion appears so symmetrical when
expressed in terms of the (anti-)BRST transformations
$s_{(a)b}$.\\

\noindent
{\bf 4 Symmetries for the Dirac fields: augmented superfield approach}\\

\noindent It is obvious from the definition and property
associated with the covariant derivative $D_\mu \psi (x) =
\partial_\mu \psi (x) + i e A_\mu (x) \psi (x)$ that an
interesting combination of the $U(1)$ gauge field $A_\mu$ and the
matter (Dirac) fields, through this derivative (i.e. $\bar \psi
(x) D_\mu \psi (x)$), is a  gauge (and, therefore, BRST) invariant
quantity. In what follows, we shall derive the exact and unique
 expressions for the nilpotent symmetry
transformations (2.2) for the matter fields by demanding the
invariance this gauge invariant quantity on the supermanifold.
This statement can be mathematically expressed by the following equation:
$$
\begin{array}{lcl}
\bar \Psi (x, \theta, \bar\theta) \;\bigl (\tilde  d + i\;e\;
\tilde A^{(1)}_{(h)} \bigr )\;
 \Psi (x, \theta, \bar\theta) &=& \bar \psi (x)\; \bigl (d + i\; e\;
 A^{(1)} \bigr )\; \psi (x),
\end{array} \eqno(4.1)
$$
where $\tilde d$ and $\tilde A^{(1)}$ are the super exterior derivative
and super 1-form connection (cf. (3.3)) on a six $(4, 2)$-dimensional
supermanifold and $d = dx^\mu \partial_\mu$ and $A^{(1)} = dx^\mu A_\mu$ are
their counterparts on the ordinary 4D Minkowskian spacetime manifold. In
particular,
 $\tilde A^{(1)}_{(h)} = dx^\mu B_\mu^{(h)} + d \theta \bar {\cal F}^{(h)}
+ d \bar\theta  {\cal F}^{(h)}$
is the expression for $\tilde A^{(1)}$ after the application
of the horizontality condition  (cf. (3.6)). The general
super expansion of the superfields $(\Psi, \bar \Psi) (x, \theta, \bar\theta)$,
corresponding to the ordinary Dirac fields $(\psi, \bar\psi)(x)$, are
taken as follows
$$
\begin{array}{lcl}
 \Psi (x, \theta, \bar\theta) &=& \psi (x)
+ i \;\theta\; \bar b_1 (x) + i \;\bar \theta \; b_2 (x)
+ i \;\theta \;\bar \theta \;f (x),
\nonumber\\
\bar \Psi (x, \theta, \bar\theta) &=& \bar \psi (x)
+ i\; \theta \;\bar b_2 (x) + i \;\bar \theta \; b_1 (x)
+ i\; \theta \;\bar \theta \;\bar f (x).
\end{array} \eqno(4.2)
$$ It is clear that, in the limit $(\theta, \bar\theta)
\rightarrow 0$, we get back the Dirac fields $(\psi, \bar\psi)$ of
the Lagrangian density (2.1). Furthermore, the number of bosonic
fields ($b_1, \bar b_1, b_2, \bar b_2)$ match with that of the
fermionic fields $(\psi, \bar \psi, f, \bar f)$ so that the above
expansion is consistent with the basic tenets of supersymmetry.


It is straightforward to see that there is only one term on the
r.h.s. of (4.1) which can be explicitly expressed as:
$dx^\mu \; [\bar \psi (x) \; (\partial_\mu + i e A_\mu (x)) \; \psi (x)]$.
However, it is evident
that on the l.h.s., we shall have the coefficients of
the differentials $dx^\mu, d \theta$ and $d \bar\theta$.
To compute explicitly these coefficients, let us first focus on the
first term of the l.h.s. of (4.1):
$$
\begin{array}{lcl}
\bar \Psi (x, \theta, \bar\theta)\; \tilde d \; \Psi (x, \theta, \bar\theta)
= \bar \Psi\; (dx^\mu\; \partial_\mu)\; \Psi
+ \bar \Psi\; (d \theta\; \partial_\theta) \;\Psi
+ \bar \Psi\; (d \bar \theta\; \partial_{\bar\theta})\; \Psi,
\end{array} \eqno(4.3)
$$
where, it can be readily checked that
$\partial_\theta \Psi = i \bar b_1 + i \bar\theta f$ and
$\partial_{\bar \theta} \Psi = i b_2 - i \theta f$. Taking
the help of the anticommuting
properties of the Grassmannian variables and their
differentials,
we obtain the following explicit expressions for the coefficients of
$d \theta$ and $d \bar\theta$ from (4.3):
$$
\begin{array}{lcl}
&&- (d \theta)\; \Bigl [\; i\; \bar\psi \; \bar b_1 - i\; \bar\theta\;
\bigl (\bar \psi\; f - i\; b_1 \;\bar b_1 \bigr ) - \theta\;
\bigl (\bar b_2\; \bar b_1 \bigr) - \theta\; \bar \theta\;
\bigl (\bar b_2\; f + \bar f \; \bar b_1 \bigr )\;\Bigr ]
\nonumber\\
&& - (d \bar\theta)\; \Bigl \{\; i\; \bar\psi \; b_2 + i\; \theta\;
\bigl (\bar \psi\; f + i\; \bar b_2\; b_2 \bigr ) - \bar\theta\;
\bigl (b_1 \; b_2 \bigr ) - \theta\; \bar\theta \;
\bigl (b_1\; f + \bar f\; b_2 \bigr)\; \Bigr \},
\end{array} \eqno(4.4)
$$
where we have exploited $d \theta\; \bar \Psi = - \bar \Psi\; d\theta$, etc.,
and the expansion of $(\Psi, \bar \Psi) (x, \theta, \bar\theta)$
from (4.2). Using, once again,
the anticommuting properties (i.e. $\theta^2 = \bar\theta^2 = 0,
\theta \bar\theta + \bar\theta \theta = 0$) of the Grassmannian variables
$\theta$ and $\bar\theta$, we obtain the following explicit expression
for the coefficient $K_\mu (x, \theta, \bar\theta)$ of $dx^\mu$ from
the first term of (4.3), namely;
$$
\begin{array}{lcl}
(dx^\mu)\; K_\mu (x, \theta, \bar\theta) \equiv (dx^\mu)\;
\Bigl [\; \bigl \{ \bar\psi \partial_\mu \psi  \bigr\} (x)
+ i\; \theta\;
L_\mu (x) + i \; \bar\theta\; M_\mu (x) + i\;\theta\;\bar \theta\; N_\mu (x)
\Bigr ],
\end{array} \eqno(4.5)
$$
where the long-hand expressions for the $L_\mu (x), M_\mu (x)$ and
$N_\mu (x)$ are:
$$
\begin{array}{lcl}
&& L_\mu (x) = \bar b_2\; \partial_\mu \; \psi - \bar \psi\;
\partial_\mu\;\bar b_1, \qquad
 M_\mu (x) = b_1\; \partial_\mu \; \psi - \bar \psi\;
\partial_\mu\;b_2, \nonumber\\
&& N_\mu (x) = \bar \psi\; \partial_\mu\; f + \bar f \; \partial_\mu \psi
+ i\; (\bar b_2\;\partial_\mu\; b_2 - b_1 \; \partial_\mu\; \bar b_1).
\end{array} \eqno(4.6)
$$

Let us concentrate on the explicit computation of the coefficients
of $dx^\mu, d \theta$ and $d \bar\theta$ that emerge from the
second term of the l.h.s. of (4.1) (i.e. $i\; e\; \bar\Psi
\;\tilde A^{(1)}_{(h)}\; \Psi)$.  This term, written in the
component fields,  has the following clear expansion $$
\begin{array}{lcl}
i\; e\; \bar \Psi (x, \theta, \bar\theta) \; \tilde
A^{(1)}_{(h)}\; \Psi (x, \theta, \bar\theta) = i\; e\; \Bigl [\bar
\Psi (dx^\mu\; B^{(h)}_\mu) \Psi + \bar \Psi\; (d \theta \bar
{\cal F}^{(h)}) \Psi + \bar \Psi\; (d \bar \theta {\cal F}^{(h)})
\Psi \Bigr ],
\end{array} \eqno(4.7)
$$
where we shall be using the neat expressions for the expansions (3.6)
obtained after the application of the horizontality condition. It is
clear that the latter two terms of (4.7) lead to the computation
of the coefficients of $d \theta$ and $d \bar\theta$. These are
as quoted below:
$$
\begin{array}{lcl}
&& - i\; e\; d \theta\; \bigl (\bar\psi  \bar C \psi \bigr )
+ e \; d \theta\; \bar \theta\; \bigl (\bar \psi \bar C
b_2 - \bar \psi B \psi + b_1 \bar C  \psi \bigr )
+ e \; d \theta\; \theta\; \bigl (\bar \psi \bar C \bar b_1 + \bar b_2
\bar C  \psi \bigr ) \nonumber\\
&& + e \; d \theta\; \theta\;\bar\theta\;
\Bigl [\bar \psi  \bar C f + \bar f \bar C \psi
+ i\; \bigl (b_1 \bar C \bar b_1 + \bar b_2 B \psi - \bar b_2 \bar C b_2
- \bar \psi B \bar b_1 \bigr )\; \Bigr ], \nonumber\\
&& - i\; e\; d \bar\theta\; \bigl (\bar\psi  C \psi \bigr )
+ e \; d \bar \theta \;\theta\; \bigl (\bar \psi C \bar b_1 + \bar \psi
 B \psi + \bar b_2 C \psi\bigr )
+ e\; d \bar \theta \;\bar \theta\;
\bigl (\bar \psi C b_2 + b_1 C \psi \bigr ),
\nonumber\\
&& + e\; d \bar\theta\; \theta\; \bar\theta\;
\Bigl [\;\bar \psi  C f + \bar f C\psi
+ i\; \bigl (b_1 C \bar b_1 + b_1 B \psi - \bar b_2 C b_2
- \bar \psi B b_2\bigr )\; \Bigr ].
\end{array} \eqno(4.8)
$$ The spacetime component (i.e. the coefficient $E_\mu (x,
\theta, \bar\theta)$ of $dx^\mu$) that emerges from the expansion
of the first term of (4.7),  is given below:
$$
\begin{array}{lcl}
(dx^\mu)\; E_\mu (x, \theta, \bar\theta) = (dx^\mu)\;
\Bigl [\; i\; e\;\bigl \{\bar\psi A_\mu \psi  \bigr\}  (x) + \theta\;
F_\mu (x) + \bar\theta\; G_\mu (x) + \;\theta\;\bar \theta\; H_\mu (x)
\Bigr ].
\end{array} \eqno(4.9)
$$
The exact expressions for $F_\mu (x), G_\mu (x)$ and $H_\mu (x)$,
in terms of component fields,  are
$$
\begin{array}{lcl}
F_\mu (x) &=& e\; \bigl [\bar \psi\;
A_\mu\;\bar b_1 - i\; \bar \psi \;\partial_\mu \bar C\; \psi
- \bar b_2\; A_\mu\; \psi \bigr ], \nonumber\\
G_\mu (x) &=& e\; \bigl (\bar \psi\;
A_\mu\;b_2 - i\; \bar \psi \;\partial_\mu C \;\psi
- b_1\; A_\mu\; \psi \bigr ), \nonumber\\
H_\mu (x) &=& - e\; \bigl [\bar \psi\;
A_\mu\;f + \bar f\; A_\mu\; \psi
- \bar \psi \;\partial_\mu \bar C \;b_2
+ \bar \psi \;\partial_\mu C \;\bar b_1
+ \bar \psi \;\partial_\mu B \;\psi \nonumber\\
&+& i\; \bar b_2\; A_\mu\; b_2
- i\; b_1\; A_\mu\; \bar b_1
- b_1\; \partial_\mu \bar C\; \psi
+ \bar b_2\; \partial_\mu C\; \psi
 \bigr ].
\end{array} \eqno(4.10)
$$
It should be noted that the above equations (4.9) and (4.10) have emerged
from the first term of (4.7). Now, we first set the Grassmannian components
(i.e. the coefficients of $d\theta$ and $d\bar\theta$) equal to zero
because there are no such type of terms for comparison on the r.h.s.
of (4.1). From equations (4.4) and (4.8),
we obtain the following terms
with $d\theta$:
$$
\begin{array}{lcl}
&&- i\; d \theta\; \bigl (\bar \psi \bar b_1
+ e \bar \psi \bar C \psi \bigr )
+ i\; d \theta \; \bar\theta\;
\bigl [\;\bar \psi f - i b_1  \bar b_1 - i\;  e \;(\bar \psi \bar C
b_2 - \bar \psi B \psi + b_1 \bar C \psi) \;\bigr ] \nonumber\\
&& + d \theta\; \theta\; \bigl [ \;\bar b_2 \bar b_1
+ e \bigl (\bar \psi \bar C \bar b_1 + \bar b_2 \bar C \psi \bigr ) \; \bigr ]
+ d \theta\; \theta\; \bar\theta\; \Bigl [\;
\bar b_2 f + \bar f \bar b_1 + e \bigl \{\;
\bar \psi  \bar C f + \bar f \bar C \psi \nonumber\\
&& + i\; \bigl (b_1 \bar C \bar b_1
+ \bar b_2 B \psi - \bar b_2 \bar C b_2
- \bar \psi B \bar b_1 \bigr ) \bigr \}\; \Bigr ].
\end{array} \eqno(4.11)
$$
Setting equal to zero the coefficients of $d \theta, d \theta (\bar\theta),
d\theta (\theta)$ and
$d \theta (\theta \bar\theta)$ independently leads to
$$
\begin{array}{lcl}
&&\bar b_1 = - e\bar C \psi, \quad
\bar \psi f - i e \bar\psi \bar C b_2 + i  e
\bar \psi B \psi = 0, \quad
\bar b_2 \bar b_1 + e\; \bigl (\bar \psi \bar C \bar b_1
+ \bar b_2 \bar C \psi \bigr ) = 0, \nonumber\\
&& \bar b_2 f + \bar f \bar b_1 + e\; \bigl \{\;
\bar \psi  \bar C f + \bar f \bar C \psi
 + i\; \bigl (b_1 \bar C \bar b_1
+ \bar b_2 B \psi - \bar b_2 \bar C b_2
- \bar \psi B \bar b_1 \bigr ) \bigr \} = 0.
\end{array} \eqno(4.12)
$$
In the second entry, we have used
$ - i b_1 \bar b_1 - i e b_1 \bar C \psi = 0$
because $\bar b_1 = - e \bar C \psi$. The analogue of (4.11),
that emerges from (4.4) and (4.8)
with the differential $d \bar\theta$, is as follows
$$
\begin{array}{lcl}
&&- i\; d \bar \theta\; \bigl (\bar \psi b_2
+ e \bar \psi C \psi \bigr )
- i\; d \bar\theta \; \theta\;
\bigl [\;\bar \psi f + i \bar b_2  b_2 + i\;  e\;(\bar \psi C
\bar b_1 + \bar \psi B \psi + \bar b_2 C \psi)\; \bigr ] \nonumber\\
&& + d\bar \theta\; \bar\theta\; \bigl [ \;b_1 b_2
+ e \bigl (\bar \psi C b_2 + b_1 C \psi \bigr ) \; \bigr ]
+ d \bar\theta\; \theta\; \bar\theta\; \Bigl [\;
b_1 f + \bar f b_2 + e \bigl \{\;
\bar \psi  C f + \bar f C \psi \nonumber\\
&& + i\; \bigl (b_1 C \bar b_1
+ b_1 B \psi - \bar b_2 C b_2
- \bar \psi B b_2 \bigr ) \bigr \}\; \Bigr ].
\end{array} \eqno(4.13)
$$
For $\bar \psi \neq 0$, we obtain the following independent relations
from the above equation:
$$
\begin{array}{lcl}
&&b_2 = - e C \psi, \quad
\bar \psi   f + i e \bar\psi C \bar b_1 + i  e
\bar \psi  B \psi = 0, \quad
b_1 b_2 + e \bigl (\bar \psi C b_2 + b_1 C \psi \bigr ) = 0,
\nonumber\\
&& b_1 f + \bar f b_2 + e\; \bigl \{\;
\bar \psi  C f + \bar f C \psi
+ i\; \bigl (b_1 C \bar b_1
+ b_1 B \psi - \bar b_2 C b_2
- \bar \psi B b_2 \bigr ) \bigr \} = 0.
\end{array} \eqno(4.14)
$$
In the above, in the second entry,
$ i \bar b_2 b_2 + i e \bar b_2 C \psi = 0$
has been exploited due to the fact that $b_2 = - e C \psi$. It can be readily
checked that the equations (4.12) and (4.14) allow the following expression
for $f$ as the solution to the second entries of both of them
$$
\begin{array}{lcl}
f = - i\; e\; \bigl (\; B + e\; \bar C\; C \;\bigr )\; \psi.
\end{array} \eqno(4.15)
$$ The substitution of all the above values for $\bar b_1, b_2$
and $f$ in (4.2) yields the following expansion of the superfield
$\Psi$ in the language of the (anti-) BRST transformations (2.2):
$$
\begin{array}{lcl}
\Psi (x, \theta, \bar \theta) = \psi (x) + \theta\; (s_{ab} \psi (x))
+ \bar \theta\; (s_b \psi (x)) + \theta\; \bar\theta\; (s_b s_{ab} \psi (x)).
\end{array} \eqno(4.16)
$$
It will be noted that, so far, the third and fourth entries of (4.12)
and (4.14) have not been exploited. We shall comment on them
a bit later (i.e. after equation (4.23)).

Now the stage is set for the discussion of the coefficients of $dx^\mu$
that emerge from (4.5) and (4.9). It is straightforward to check that
the coefficients of the pure $dx^\mu$ from l.h.s. and r.h.s. do
match. Furthermore, the coefficient of $dx^\mu\; \theta$  is
ought to be zero because there is no such term on the r.h.s..
The exact expression for such an equality is as follows
$$
\begin{array}{lcl}
e\; \bigl (\bar \psi\; A_\mu\; \bar b_1 - i\; \bar \psi\; \partial_\mu \bar C
\;\psi - \bar b_2\; A_\mu \; \psi \bigr )
+ i\; \bigl (\bar b_2\; \partial_\mu \psi - \bar \psi\; \partial_\mu \bar b_1
\bigr ) = 0.
\end{array} \eqno(4.17)
$$
Exploiting the values of $\bar b_1$ and $b_2$ from (4.12) and (4.14), the
above equation leads to the following useful equation for the unknown
local parameter field $\bar b_2 (x)$:
$$
\begin{array}{lcl}
i\; \bigl (\; \bar b_2 + e \; \bar \psi\; \bar C\; \bigr)\;
 \bigl (\; \partial_\mu \psi + i e A_\mu \psi\; \bigr)\; = 0.
\end{array} \eqno(4.18)
$$
This yields the value for $\bar b_2$ to be $ - e \bar \psi \bar C$
(i.e. $\bar b_2 = - e \bar \psi \bar C$). It is clear that
$D_\mu \psi = \partial_\mu \psi + i e A_\mu \psi \neq 0$ for an interacting
$U(1)$ gauge theory because the interaction term
$- e \bar \psi \gamma^\mu A_\mu \psi$
is hidden in the covariant derivative in the sense
that it emerges from $i \bar\psi \gamma^\mu D_\mu \psi$. Setting
equal to zero the coefficient of $dx^\mu \; \bar\theta$, we obtain
$$
\begin{array}{lcl}
e\; \bigl (\bar \psi\; A_\mu\; b_2 - i\; \bar \psi\; \partial_\mu C
\;\psi - b_1\; A_\mu \; \psi \bigr )
+ i\; \bigl (b_1\; \partial_\mu \psi - \bar \psi\; \partial_\mu b_2
\bigr ) = 0.
\end{array} \eqno(4.19)
$$
Substituting the value of $b_2$ (i.e. $ b_2 = - e\; C\; \psi$), we
obtain the following relation for an unknown local parameter
component field $b_1 (x)$ of expansion (4.2):
$$
\begin{array}{lcl}
i\; \bigl (\; b_1 + e \; \bar \psi\; C\; \bigr)\;
 \bigl (\; \partial_\mu \psi + i e A_\mu \psi\; \bigr)\; = 0.
\end{array} \eqno(4.20)
$$
The above equation produces the value of $b_1$  as $- e\; \bar\psi\; C$
in a unique fashion.
Ultimately, we now focus on the computation of the coefficient of
$dx^\mu\;\theta\;\bar\theta$ which will naturally
be set equal to zero because there is no such term on the r.h.s.
Mathematically, the precise expression, for the above
statement of equality is as follows
$$
\begin{array}{lcl}
&-& e\; \Bigl [\;\bar \psi\; A_\mu\; f + \bar f\; A_\mu\; \psi
- \bar\psi\; \partial_\mu \bar C\; b_2
+ \bar\psi\; \partial_\mu C\; \bar b_1
+ \bar\psi\; \partial_\mu B\; \psi\nonumber\\
&+& i\; \bar b_2 \; A_\mu\; b_2
- i\; b_1 \; A_\mu\; \bar b_1 - b_1 \;\partial_\mu \bar C\; \psi
+ \bar b_2\; \partial_\mu C\; \psi\; \Bigr ] \nonumber\\
&+& i\; \Bigl [\;\bar \psi\; \partial_\mu f + \bar f\; \partial_\mu \psi
+ i\; \bigl (\bar b_2\; \partial_\mu b_2 - b_1\;\partial_\mu \bar b_1
\bigr )\; \Bigr ] = 0.
\end{array} \eqno(4.21)
$$
Plugging in the values $b_1 = - e \bar \psi C, \bar b_1 = - e \bar C \psi,
b_2 = - e C \psi, \bar b_2 = - e \bar \psi \bar C,
f = - i e (B + e \bar C C)\; \psi$, we obtain the following relationship
for the unknown local parameter field $\bar f (x)$
$$
\begin{array}{lcl}
i\; \bigl (\bar f - i \; e\; \bar \psi\; B + i\; e^2\; \bar \psi\;
\bar C \; C \bigr)\;
 \bigl (\; \partial_\mu \psi + i e A_\mu \psi\; \bigr)\; = 0.
\end{array} \eqno(4.22)
$$
The above relation yields the expression for $\bar f(x)$ in terms of the
fields of the (anti-)BRST invariant Lagrangian density (2.1).
Together, all the local (unknown) secondary component fields, in the
expansion (4.2) of the superfields $(\Psi, \bar\Psi) (x, \theta, \bar\theta)$,
are as follows
$$
\begin{array}{lcl}
&& b_1 = - e \bar \psi C, \qquad b_2 = - e C \psi,
\qquad \bar b_1 = - e \bar C \psi, \qquad \bar b_2 = - e \bar \psi \bar C,
\nonumber\\
&& f = - i e\; [\; B + e \bar C C\; ]\; \psi,
\qquad \bar f = + i e\; \bar \psi\; [\; B + e C \bar C \;].
\end{array} \eqno(4.23)
$$
It is worthwhile to mention that, exactly the same expressions
as quoted above,
were obtained in our earlier work [11] where the invariance of the
conserved matter current on the supermanifold was imposed. However,
the above solutions in [11] were not mathematically
{\it unique}. In our present endeavour, we have been able
to show the uniqueness and exactness of the solutions (4.23).
Furthermore, the solutions (4.23) do {\it satisfy} all the conditions of
(4.12) and (4.14) which have appeared as the third and fourth entries.
With the values from (4.23),
the super expansion of the superfield $\bar \Psi (x, \theta, \bar\theta)$,
in the language of the (anti-) BRST transformations (2.2), is as
illustrated below
$$
\begin{array}{lcl}
\bar \Psi (x, \theta, \bar \theta) = \bar \psi (x)
+ \theta\; (s_{ab} \bar\psi (x))
+ \bar \theta\; (s_b \bar\psi (x))
+ \theta\; \bar\theta\; (s_b s_{ab} \bar\psi (x)).
\end{array} \eqno(4.24)
$$
The above expansion, in terms of $s_{(a)b}$, bears exactly the same
appearance as its counterpart in (4.16) where the expansion for the superfield
$\Psi (x, \theta, \bar\theta)$ has been given.

\noindent
{\bf 5 Conclusions}\\

\noindent
The long-standing problem of the derivation of the nilpotent
(anti-) BRST symmetry transformations for the matter (e.g.  Dirac) fields
of an interacting gauge theory (e.g. QED), in the
framework of superfield formalism
\footnote{ In the known literature on
the usual superfield formulation, only the nilpotent BRST-type symmetry
transformations for the gauge- and (anti-) ghost fields have been
derived without any comment on such type of transformations
associated with the matter fields of  an interacting
gauge theory [1-5,10]. However, in our recent works on the augmented
superfield formalism [11-16], this problem has been addressed.},
has been resolved  uniquely in our present endeavour. We have invoked
an additional  gauge invariant restriction (cf. (4.1)), besides the
usual horizontality condition (cf. (3.2)), on the supermanifold to obtain
the off-shell nilpotent symmetry transformations for the Dirac  fields
of QED.

It is very interesting and gratifying that {\it both} the
restrictions on the supermanifold are complementary and consistent
with each-other and, more importantly, they are intertwined in the
sense that they owe their origin to the nilpotent (super) exterior
derivatives $(\tilde d)d$ and 1-form (super) connection $(\tilde
A^{(1)})A^{(1)}$. The present extended version of the usual
superfield formalism, which leads to the derivation of
mathematically exact expressions for the off-shell nilpotent
(anti-) BRST symmetry transformations associated with {\it all}
the fields of QED, has been christened as the {\it augmented}
superfield approach to BRST formalism.

It is worthwhile to note that the horizontality condition (cf.
(3.2)) on the supermanifold (that leads to the derivation of exact
nilpotent symmetry transformations for the gauge- and (anti-)
ghost fields) is precisely a gauge covariant statement because,
for the non-Abelian $SU(N)$ gauge theory, the 2-form curvature
$F^{(2)}$ transforms as $F^{(2)} \to (F^{(2)})^\prime = U F^{(2)}
U^{-1}$ (with $U \in SU(N)$). It is another matter that it becomes
a gauge invariant statement (i.e. $F^{(2)} \to (F^{(2)})^\prime =
U F^{(2)} U^{-1} = F^{(2)}$) for our present case of an
interacting Abelian  $U(1)$ gauge theory. In contrast, the
additional restriction (4.1), invoked on the supermanifold, is
primarily a gauge invariant statement. In fact, its gauge
covariant version on the supermanifold leads to absurd results
(even for the simplest case of an interacting Abelian $U(1)$ gauge
theory), as can be seen explicitly in the Appendix.

Our present theoretical arsenal of the augmented superfield
formalism has already  been exploited [18,19] for the derivation
of the exact and unique nilpotent symmetry transformations for (i)
the complex scalar fields in interaction with the $U(1)$ gauge
field (see, e.g. [11,12] for earlier works), and (ii) the Dirac
fields in interaction with the $SU(N)$ non-Abelian gauge field
(see, e.g., [14] for our earlier work). As is evident from our
discussions that $B_\mu, {\cal F}, \bar {\cal F}$ form the vector
multiplet of the 1-form superfield $\tilde A^{(1)} = d Z^M \tilde
A_M$. One of the most intriguing question, in this context, is to
find out some multiplet of superfield that can accommodate the
spinor superfields $\Psi (x,\theta,\bar\theta)$ and $\bar \Psi (x,
\theta, \bar\theta)$. So far, we have not been able to get this
multiplet. It would be interesting to find the answer to this
question. These are some of the immediate and urgent issues that
are under investigation at the moment and our results will be
reported in our future publications [20].\\

\noindent
{\bf Acknowledgements}\\

\noindent
Fruitful conversations with L. Bonora (SISSA, Italy), K. S. Narain
(AS-ICTP, Italy) and M. Tonin (Padova, Italy) are gratefully acknowledged.
The warm hospitality extended at the AS-ICTP and Physics Dept.
(Padova University) is thankfully acknowledged, too.\\

\vskip 0.2cm

\begin{center}
{\bf {\Large Appendix}}
\end{center}

\vskip 0.2cm

\noindent
Let us begin with the gauge {\it covariant} version of (4.1),
namely;
$$
\begin{array}{lcl}
\bigl (\tilde d + i e \tilde A^{(1)}_{(h)} \bigr ) \; \Psi (x,\theta,\bar\theta)
= \bigl ( d + i e A^{(1)} \bigr ) \; \psi (x),
\end{array}\eqno(A.1)
$$ where the symbols carry their usual meaning, as discussed
earlier. It is clear that the r.h.s. of the above equation (i.e.
$dx^\mu (\partial_\mu + i e A_\mu) \psi (x)$) contains a single
differential $dx^\mu$. However, the l.h.s. would yield the
coefficients of differentials $dx^\mu, d \theta$ and $d
\bar\theta$. In fact, the l.h.s. consists of $\tilde d \Psi$ and
$i e \tilde A^{(1)}_{(h)} \Psi$. The former can be written as $$
\begin{array}{lcl}
\tilde d\; \Psi  = d x^\mu (\partial_\mu \psi + i \theta
\partial_\mu \bar b_1 + i \bar\theta \partial_\mu b_2 + i \theta
\bar\theta \partial_\mu f) + i d \theta ( \bar b_1 + \bar\theta f)
+ i d \bar \theta (b_2 -  \theta f),
\end{array}\eqno(A.2)
$$
and the latter term can be explicitly expressed as
$$
\begin{array}{lcl}
i e \tilde A^{(1)}_{(h)} \Psi (x, \theta,\bar\theta)
= i e [\; dx^\mu B_\mu^{(h)} + d \theta \bar {\cal F}^{(h)} + d \bar\theta
{\cal F}^{(h)} \;]\; \Psi (x,\theta,\bar\theta).
\end{array}\eqno(A.3)
$$
The latter two terms of the above expression yield the following
coefficients of $d \theta$ and $d \bar\theta$
$$
\begin{array}{lcl}
i e d \theta (\bar C \psi) + e d\theta (\theta) (\bar C \bar b_1)
+ e d \theta (\bar \theta) [ \bar C b_2 - B \psi] - e d\theta
(\theta \bar\theta) [\bar C f - i B \bar b_1],
\end{array}\eqno(A.4)
$$
$$
\begin{array}{lcl}
i e d \bar\theta (C \psi) + e d \bar\theta (\theta) [ C \bar b_1 + B
\psi ] + e d \bar\theta (\bar\theta) (C b_2) + e d \bar\theta
(\theta \bar\theta) [ C f - i B b_2 ].
\end{array}\eqno(A.5)
$$
It is straightforward to note that the above coefficients would
not emerge from the r.h.s. Thus, these coefficients would be
set equal to zero. Equating the coefficients of $d\theta$,
$d \theta (\theta), d \theta (\bar\theta)$ and $d\theta (\theta
\bar\theta)$ equal to zero, we obtain the following conditions
$$
\begin{array}{lcl}
\bar b_1 = - e \bar C \psi, \qquad \bar C \bar b_1 = 0, \qquad
f = - i e (B\psi - \bar C b_2),
\qquad \bar C f - i B \bar b_1 = 0.
\end{array}\eqno(A.6)
$$
It should be noted that, in the above computation, exactly similar
type of terms have been collected from (A.2) and (A.4). It is
obvious that the second and fourth conditions are satisfied
if we take into account the value of $\bar b_1, f$ and
exploit the condition $\bar C^2 = 0$. Similarly,
we collect the terms of similar kinds from (A.2) and (A.5) and
set the coefficients of $d\bar\theta, d \bar\theta (\theta),
d\bar\theta (\bar\theta)$ and $d \bar\theta (\theta\bar\theta)$
equal to zero separately and independently. These lead to
the following conditions
$$
\begin{array}{lcl}
b_2 = - e C \psi, \qquad f = - i e (B \psi + C \bar b_1),
\qquad C b_2 = 0, \qquad C f - i B b_2 = 0.
\end{array}\eqno(A.7)
$$
It is evident that, in the above, the third and fourth conditions
are satisfied. Finally, exploiting the conditions in (A.6)
and (A.7), we obtain the following
$$
\begin{array}{lcl}
\bar b_1 = - e \bar C \psi, \qquad b_2 = - e C \psi, \qquad f = -
i e (B + e \bar C C) \psi.
\end{array}\eqno(A.8)
$$
It is worth emphasizing that the above results are also
obtained from the gauge invariant condition (4.1) when we
set equal to zero the coefficients of $d\theta$
and $d\bar\theta$.

The key difference between the gauge invariant condition (4.1)
and the gauge covariant condition (A.1) are found to be
contained in the coefficients of $dx^\mu$. To make this
statement more transparent, we expand the first term
(i.e. $i e\; d x^\mu \;B_\mu^{(h)}$) of (A.3) as follows
$$
\begin{array}{lcl}
i e dx^\mu \Bigl [ A_\mu \psi + i \theta (A_\mu \bar b_1 - i
\partial_\mu \bar C \psi) + i \bar\theta (A_\mu b_2 - i
\partial_\mu C \psi) + i \theta \bar\theta Q_\mu \Bigr ],
\end{array}\eqno(A.9)
$$
where the explicit expression for the quantity $Q_\mu$ is
$$
\begin{array}{lcl}
Q_\mu = A_\mu f + \partial_\mu B \psi + \partial_\mu C \bar b_1
- \partial_\mu \bar C b_2.
\end{array}\eqno(A.10)
$$
A careful observation of the equations (A.2) and (A.9)
demonstrates that
there are coefficients of $dx^\mu, dx^\mu (\theta),
dx^\mu (\bar\theta)$ and $dx^\mu (\theta\bar\theta)$.
It is straightforward to note that coefficient of
pure $dx^\mu$ from the l.h.s. does match with the one
that emerges from the r.h.s. The coefficients of
$dx^\mu (\theta), dx^\mu (\bar\theta)$ and $dx^\mu
(\theta\bar\theta)$ are listed below
$$
\begin{array}{lcl}
&& dx^\mu\; (\theta) \; \Bigl [ \; i \partial_\mu \bar b_1
- e A_\mu \bar b_1 + i e \partial_\mu \bar C \psi \;\Bigr ],
\nonumber\\ && dx^\mu \;(\bar\theta)\; \Bigl [\; i \partial_\mu b_2
- e A_\mu b_2 + i e \partial_\mu C \psi \;\Bigr ], \nonumber\\
&& dx^\mu (\theta\bar\theta) \; \Bigl [\; i \partial_\mu f - e A_\mu
f + e \partial_\mu \bar C b_2 - e \partial_\mu C \bar b_1 - e
\partial_\mu B \psi \;\Bigr ].
\end{array}\eqno(A.11)
$$
As is evident, these coefficients are set to be zero to have
the conformity with the gauge covariant condition in (A.1). The
substitution of the values from (A.8) into the above conditions
leads to the following restrictions
$$
\begin{array}{lcl}
- i e \bar C D_\mu \psi = 0, \qquad - i e C D_\mu \psi = 0,
\qquad e (B - e C \bar C) D_\mu \psi = 0.
\end{array}\eqno(A.12)
$$ The above restrictions do not lead to any physically
interesting solutions because they imply $D_\mu \psi = 0$ for $C
\neq 0, \bar C \neq 0, B \neq e C \bar C$. However, for an
interacting Abelian $U(1)$ gauge theory, this condition is absurd.
The other choices, for instance, the conditions: $C = 0, \bar C =
0$ and $B = e C \bar C$ (for $D_\mu \psi \neq 0$), are also not
acceptable. Thus, we conclude  that the gauge covariant condition
(A.1) does not lead to exact derivation of the nilpotent symmetry
transformations for the matter fields in QED. In contrast, the
gauge invariant restriction (4.1) does lead to exact derivations.

\end{document}